\documentclass[]{spie}  

\usepackage{amsmath,amsfonts,amssymb}
\usepackage{graphicx}
\usepackage{threeparttable}
\usepackage[colorlinks=true, allcolors=blue]{hyperref}
\usepackage{here}

\title{Development of the New Multi-Beam Receiver and Telescope Control System for NASCO}

\author[a,b]{Atsushi Nishimura}
\author[b]{Akio Ohama}
\author[b,c]{Kimihiro Kimura}

\author[b]{Daichi Tsutsumi}
\author[b]{Yudai Matsue}
\author[b]{Rin Yamada}
\author[b]{Mariko Sakamoto}
\author[b]{Kenta Matsunaga}

\author[a]{Yutaka Hasegawa}
\author[a]{Taisei Minami}
\author[a]{Takeru Matsumoto}

\author[b]{Kazuki Shiotani}
\author[b]{So Okuda}
\author[b]{Kakeru Fujishiro}
\author[b]{Keisuke Sakasai}
\author[b]{Masahiro Suzuki}
\author[b]{Shun Saeki}
\author[b]{Kouki Satani}
\author[b]{Kousuke Urushihara}
\author[b]{Chiharu Kato}
\author[b]{Takashi Kondo}
\author[b]{Kazuki Okawa}
\author[b]{Daiki Kurita}
\author[b]{Tetsuta Inaba}
\author[b]{Shohei Maruyama}
\author[b]{Masako Koga}
\author[b]{Kenya Noda}
\author[b]{Mikito Kohno}
\author[b]{Hiroaki Iwamura}
\author[b]{Yuki Hyoto}

\author[b]{Yuichi Hori}
\author[b]{Kaoru Nishikawa}
\author[b]{Takeru Nishioka}
\author[b]{Thoqin Pang}

\author[b,d]{Hidetoshi Sano}
\author[b,e]{Rei Enokiya}
\author[b]{Satoshi Yoshiike}
\author[a,b]{Shinji Fujita}
\author[b]{Katsuhiro Hayashi}
\author[b]{Kazufumi Torii}
\author[b]{Takahiro Hayakawa}
\author[b]{Akio Taniguchi}

\author[b]{Kisetsu Tsuge}
\author[b]{Yumiko Yamane}
\author[b]{Yusuke Hattori}
\author[b]{Takahiro Ohno}

\author[a]{Shota Ueda}
\author[a]{Sho Masui}
\author[a]{Yasumasa Yamasaki}
\author[a]{Hiroshi Kondo}

\author[f]{Kazuji Suzuki}
\author[g]{Kazuhiro Kobayashi}
\author[d]{Yasunori Fujii}
\author[b]{Yumi Fujii}
\author[d,h]{Tetsuhiro Minamidani}
\author[d]{Takeshi Okuda}
\author[b]{Hiroaki Yamamoto}
\author[b]{Kengo Tachihara}
\author[a]{Toshikazu Onishi}
\author[f]{Akira Mizuno}
\author[a]{Hideo Ogawa}
\author[b]{Yasuo Fukui}

\affil[a]{Department of Physical Science, Graduate School of Science, Osaka Prefecture University, 1-1 Gakuen-cho, Naka-ku, Sakai, Osaka 599-8531, Japan}
\affil[b]{Department of Physics, Nagoya University, Chikusa-ku, Nagoya 464-8602, Japan}
\affil[c]{Institute of Space and Astronautical Science, Japan Aerospace Exploration Agency, 3-1-1 Yoshinodai, Chuo-ku, Sagamihara, Kanagawa 252-5210, Japan}
\affil[d]{National Astronomical Observatory of Japan (NAOJ), National Institutes of Natural Sciences (NINS) 2-21-1, Osawa, Mitaka, Tokyo 181-8588, JAPAN}
\affil[e]{Department of Physics, Faculty of Science and Technology, Keio University, 3-14-1 Hiyoshi, Kohoku-ku, Yokohama, Kanagawa 223-8522, Japan}
\affil[f]{Institute for Space-Earth Environmental Research (ISEE), Nagoya University, Furo-cho, Chikusa-ku, Nagoya 464-8601 Japan}
\affil[g]{Instrument Development Center, Nagoya University, Chikusa-ku, Nagoya 464-8602, Japan}
\affil[h]{Department of Astronomical Science, The Graduate University for Advanced Studies, SOKENDAI 2-21-1, Osawa, Mitaka, Tokyo 181-8588, JAPAN}

\authorinfo{Further author information: (Send correspondence to A.N.)\\A.N.: E-mail: nishimura@p.s.osakafu-u.ac.jp}

\begin{document}
\maketitle

\begin{abstract}
We report the current status of the NASCO (NAnten2 Super CO survey as legacy) project which aims to provide all-sky CO data cube of southern hemisphere using the NANTEN2 4-m submillimeter telescope installed at the Atacama Desert through developing a new multi-beam receiver and a new telescope control system. The receiver consists of 5 beams. The four beams, located at the four corners of a square with the beam separation of 720$''$, are installed with a 100 GHz band SIS receiver having 2-polarization sideband-separation filter. The other beam, located at the optical axis, is installed with a 200 GHz band SIS receiver having 2-polarization sideband-separation filter. The cooled component is modularized for each beam, and cooled mirrors are used. The IF bandwidths are 8 and 4 GHz for 100 and 200 GHz bands, respectively. Using XFFTS spectrometers with a bandwidth of 2 GHz, the lines of $^{12}$CO, $^{13}$CO, and C$^{18}$O of $J$=1$-$0 or $J$=2$-$1 can be observed simultaneously for each beam. The control system is reconstructed on the ROS architecture, which is an open source framework for robot control, to enable a flexible observation mode and to handle a large amount of data. The framework is commonly used and maintained in a robotic field, and thereby reliability, flexibility, expandability, and efficiency in development are improved as compared with the system previously used. The receiver and control system are installed on the NANTEN2 telescope in December 2019, and its commissioning and science verification are on-going. We are planning to start science operation in early 2021. 
\end{abstract}


\section{INTRODUCTION}
\label{sec:intro}

\begin{figure} [b]
\begin{center}
\begin{tabular}{c} 
\includegraphics[width=12cm, bb=0 0 4899 2821]{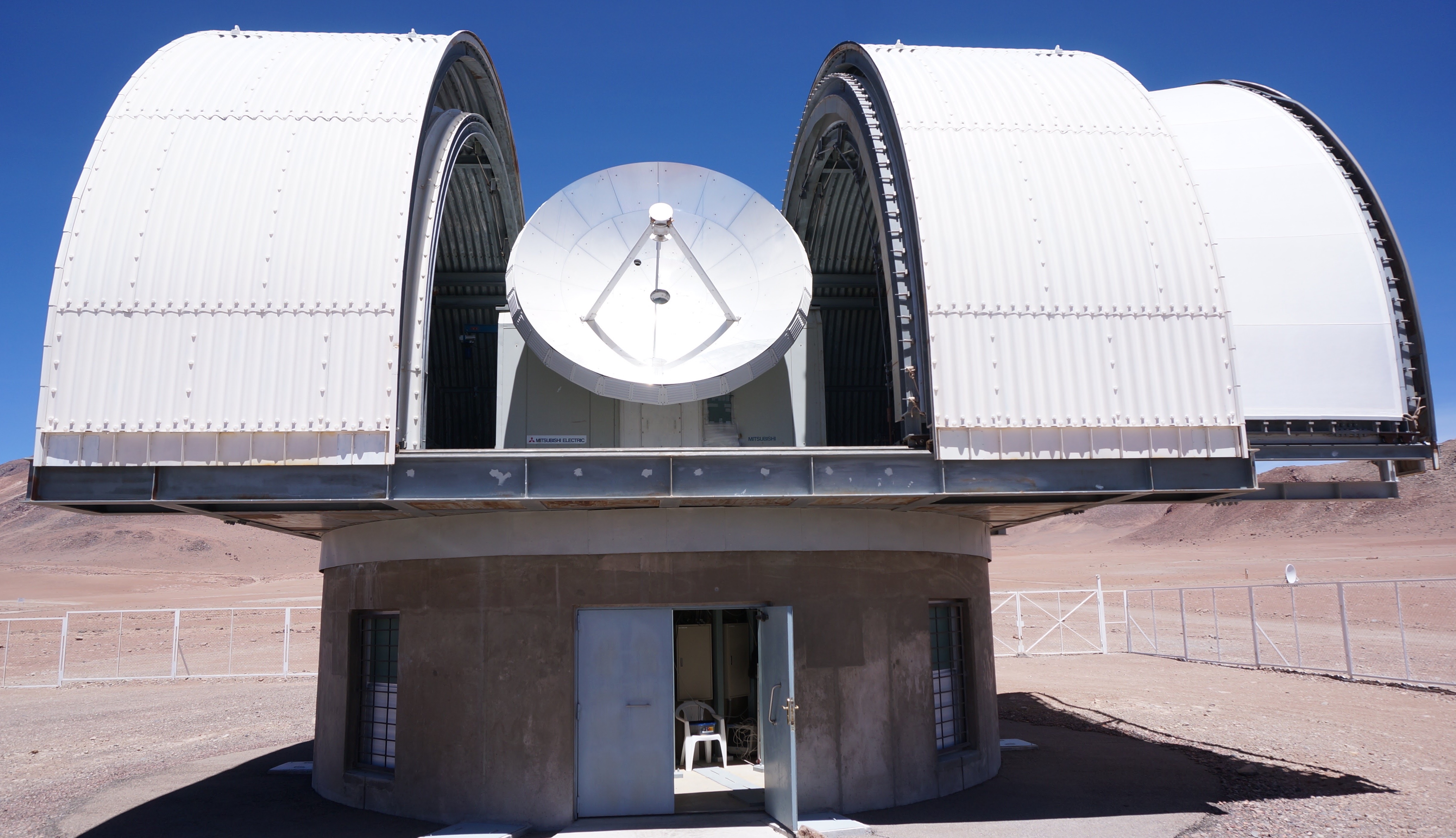}
\end{tabular}
\end{center}
\caption[example] 
{ \label{fig:1} 
NANTEN2 telescope installed at the Atacama Desert (alt. 4800 m).
}
\end{figure} 

The interstellar medium (ISM) is one of the most essential elements consisting the Universe and plays an important role in the galaxy evolution through the cycle of baryons between stars and gas.
Because of the importance of neutral gas which accounts for most of the mass of the ISM and is directly connected to the process of star formation, a number of survey projects targeting $^{12}$CO $J=$1--0 line, which is believed as the best tracer of mass of the cold neutral ISM, were conducted \cite{2001ApJ...547..792D, 2004ASPC..317...59M, 2006ApJS..163..145J, 2013PASA...30...44B, 2017PASJ...69...78U}.
The data provided by those projects were essential to achieve the current-day understanding of the ISM and were used as a guide for the further detailed observations performed by larger telescopes like ALMA.
However, the coverage and angular resolution of the existing surveys were limited because the emission from cold ISM is very weak, and, therefore, observations require relatively long integration time for each pixel.

In order to improve the observation capability, the system noise temperature should be decreased.
Recent progress in the development of mm and sub-mm receivers for the ALMA is outstanding, and the receivers having a wide bandwidth of the observed frequency adopting the 2SB technique are realized as well as the receiver noise temperatures which are close to the quantum limitation.
Some of those technologies are applied for the receivers of single dish telescopes, and the powerful receiver systems, such as T100\cite{2008PASJ...60..435N} and TZ\cite{2013PASP..125..252N} for Nobeyama 45-m telescope and the 230 GHz receiver for the 1.85-m telescope\cite{2013PASJ...65...78O, 2020SPIE.nishimura.1p85}, were developed.
For further improvement of the survey capability, the FOREST receiver\cite{2016SPIE.9914E..1ZM} was developed which is equipped with eight 2SB SIS mixers and were used to observe 4 beams with dual polarizations simultaneously, and 10 deg$^{2}$ scale mapping observations with an angular resolution of $\sim 15''$ became available\cite{2017PASJ...69...78U, 2018PASJ...70S..42N, 2019PASJ...71S...3N, 2019PASJ...71S..14S, 2019ApJ...883..156T, 2019ApJ...872...49F, 2019PASJ..tmp...46F, 2020PASJ..tmp..241N}.

In this situation, we have developed a new 115 and 230 GHz bands multi-beams receiver and installed it to the NANTEN2 4-m telescope for the purpose of the observation of all sky ($\sim 10^{4}$ deg$^{2}$) accessible from the Atacama Desert with an angular resolution of 180$''$.
We have also developed a new telescope control system to handle the large data rate and to operate faster scan on NANTEN2.
In this paper, we describe the project overview, the developed receiver and software system, and the results of science verifications and test observations using the new system.

\section{NASCO PROJECT}
\label{sec:nasco}

\subsection{NANTEN2 Telescope}

The NANTEN2 telescope (Fig. \ref{fig:1}) is installed at the Atacama Desert in Chile (alt. 4800 m) and have been operated with the two receivers since 2005: a 230 GHz receiver and a 490/810 GHz multi-beam receiver SMART\cite{2003SPIE.4855..322G}.
The 4 m main reflector consists of 33 mirrors, the typical surface accuracy of which is measured to be $<15$ $\mu$m.
The radio signals collected by the main reflector is guided to the receiver room by a Nasmyth type optics.
The mount type of the telescope is an altazimuth, and receivers are installed in the cabin which moves with the azimuth rotation.

\subsection{Science Motivations}

The ISM is an essential ingredient in understanding galaxies, and the accuracy of the ISM mass estimate affects all kinds of studies on astrophysics (e.g., galactic dynamics, formation of molecular gas, star formation, and the origin of cosmic rays).
However, currently existing $^{12}$CO $J=$1--0 data are limited in survey area or angular resolution (see for a review Ref. \citenum{2015ARA&A..53..583H}). 
The largest CO map ever observed is obtained by two 1.2-m telescopes\cite{2001ApJ...547..792D}, and its coverage is 9000 deg$^{2}$ (22 \% of the all sky), whereas the angular resolution of the survey is $>540''$, that is not high enough to study star formation which requires to resolve giant molecular clouds (GMCs) with a scale of pc in the Galaxy.
The second largest map is obtained by the NANTEN telescope\cite{2004ASPC..317...59M}.
The coverage of 4900 deg$^{2}$ (12 \% of the all sky) is observed with a beam size of 180$''$, whereas the observations were conducted in the position switching mode with a grid spacing coarser than the beam size.

In the last decade, we have a strong demand for expanding the coverage of a CO survey because there are already all sky datasets taken at the other wavelengths, which include the sub-mm data with Planck\cite{2014A&A...571A..11P} and gamma ray data with the Fermi satellite\cite{2009ApJ...697.1071A}.
A comparative study among these wavelengths should be crucial in promoting further the ISM studies\cite{2014ApJ...796...59F, 2019ApJ...884..130H}.
The NASCO project aims to observe the $^{12}$CO $J=$1--0 line for 70 \% of all sky observable from the Atacama Desert using NANTEN2. 
Fully sampled data with an angular resolution of 200$''$ will be obtained by the On-the-Fly (OTF) observation mode.
In addition to the $^{12}$CO $J=$1--0, $^{12}$CO $J=$2--1 line is also observed by the NASCO receiver.
By combining the two lines, the physical properties (e.g., temperature and volume density) of GMCs will be obtained without losing the whole CO emission\cite{2015ApJS..216...18N}.

 \begin{figure} [b]
\begin{center}
\begin{tabular}{c} 
\includegraphics[width=13cm, bb=0 0 1028 479]{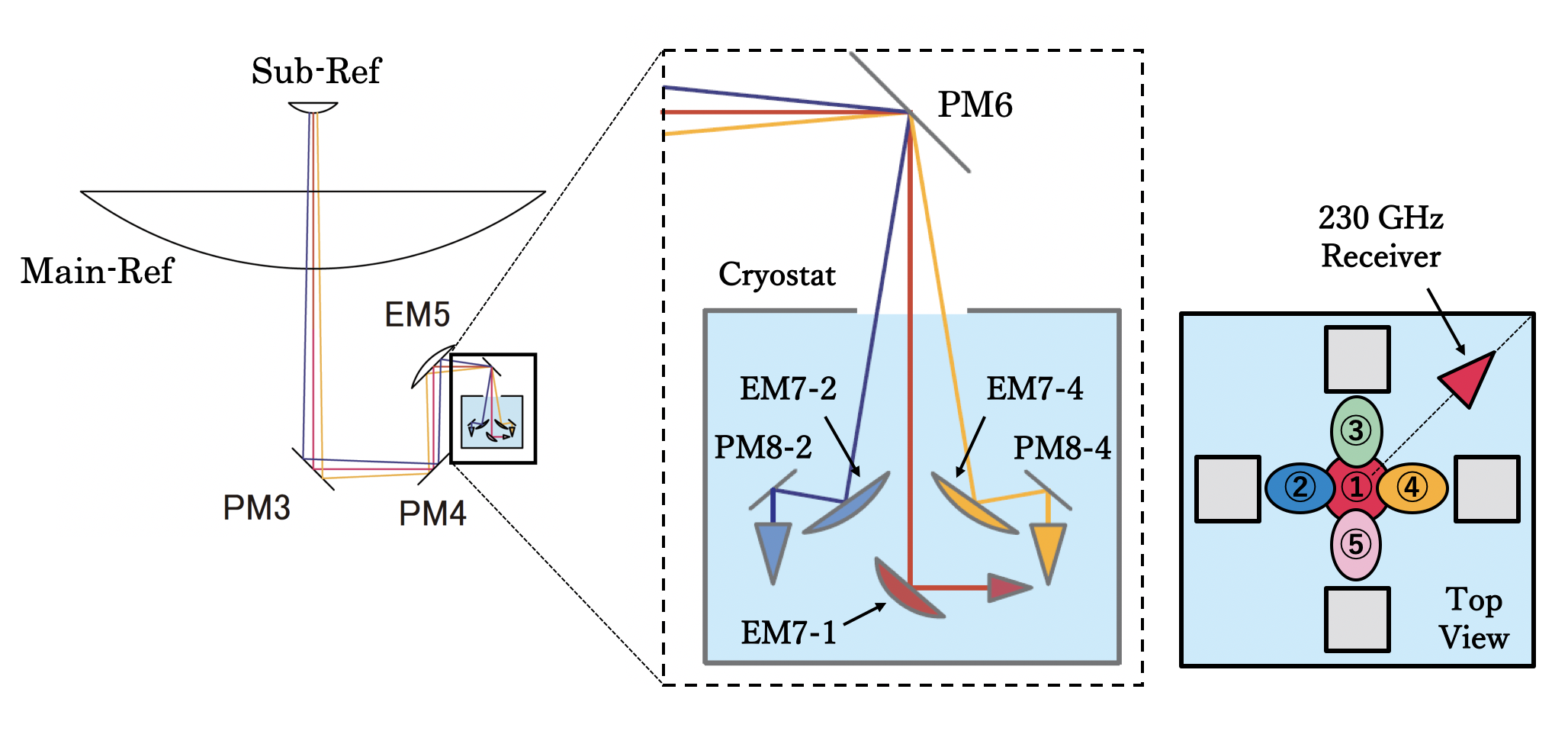}
\end{tabular}
\end{center}
\caption[example] 
{ \label{fig:optics} 
Schematic of the NASCO 110/230 GHz band 5 beam splitting optics.
}
\end{figure} 

\subsection{Requirements for the Receiver and Control System}

The noise level requirement of the NASCO project is $T_{\rm rms}=0.9$ K in $T_{\rm mb}$ scale with an angular resolution of 200$''$ and a velocity resolution of 0.2 km s$^{-1}$.
In order to achieve this noise level, 4 beams of a dual polarizations receiver with a system noise temperature of $T_{\rm sys}=200$ K is required.
Using this receiver, the noise level of $T_{\rm rms}=0.85$ K can be achieved by the observations with an integration time of 0.1 s.
Each of the XFFTS\cite{2012A&A...542L...3K} spectrometers, which is used as the back-end of the NASCO Rx, outputs the power spectrum by 32768-point double floating values (64 bit), and the NASCO Rx output 20 IF signals, thus total data rate of the NASCO project is estimated to be 420 Mbps (52 MB/s).
For the control system, capability to handle 420 Mbps data and the accurate telescope control during the fast scan of OTF observations are required.

 \begin{figure} [b]
\begin{center}
\begin{tabular}{c} 
\includegraphics[width=15cm, bb=0 0 1317 960]{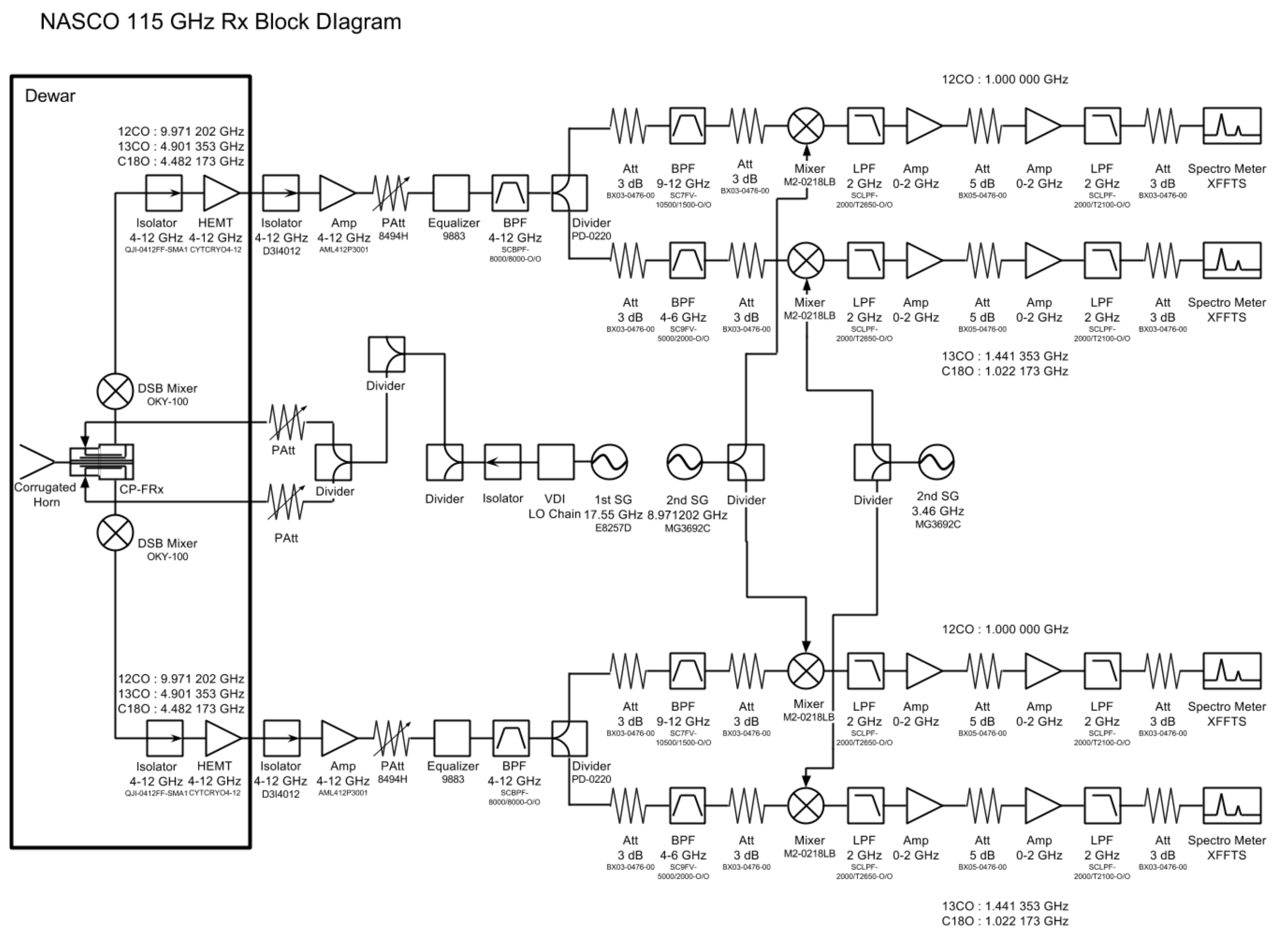}
\end{tabular}
\end{center}
\caption[example] 
{ \label{fig:ifbd100} 
Schematic block diagram of the NASCO 110 GHz band  receiver system.
}
\end{figure} 

\begin{figure} [t]
\begin{center}
\begin{tabular}{c} 
\includegraphics[width=13cm, bb=0 0 529 180]{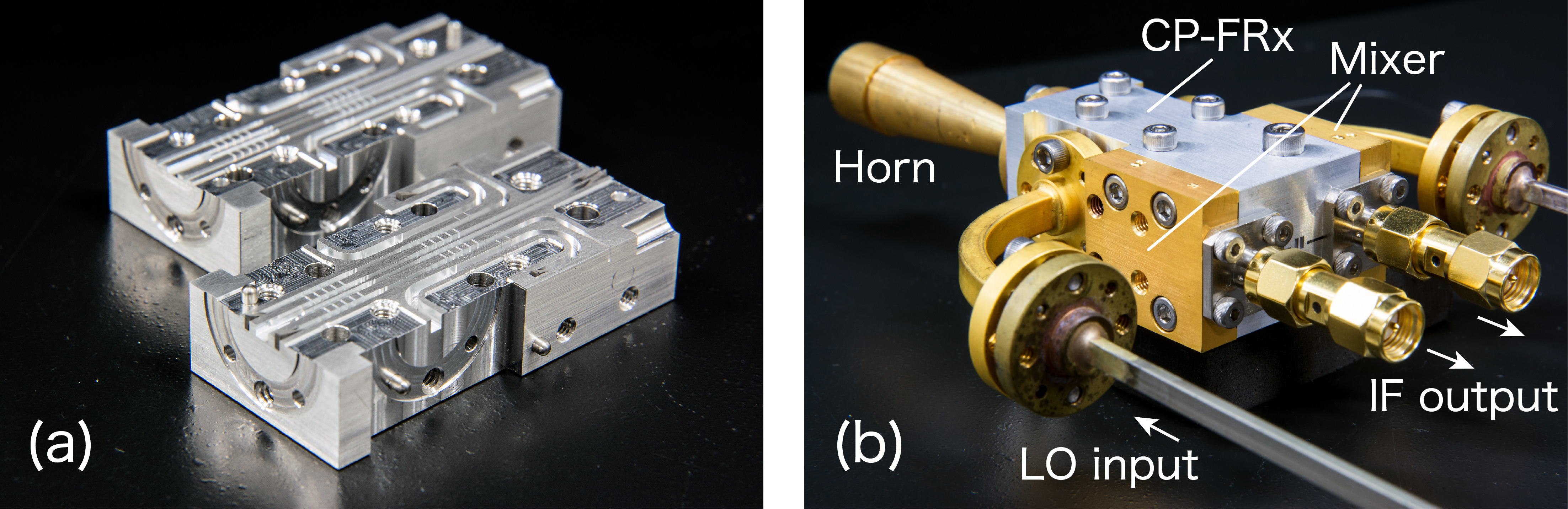}
\end{tabular}
\end{center}
\caption[example] 
{ \label{fig:rx100} 
(a) Photograph of the NASCO 110 GHz band CP-FRx Unit and
(b) the 110 GHz band SIS-Mixer receiver system front end unit. 
}
\end{figure}

\section{OPTICS AND RECEIVER: NASCO RX}
\label{sec:nascorx} 

\subsection{Optics Designing}
\label{sec:optics}
The NASCO receiver system (hereafter NASCO Rx) is a 5 beams receiver system: four beams for the 109-116 GHz band targeting $^{12}$CO, $^{13}$CO, and C$^{18}$O $J=$1--0 lines (named as beam 2--5), and one beam for the 220/230 GHz band targeting $^{12}$CO, $^{13}$CO, and C$^{18}$O $J=$2--1 lines (named as beam 1).
Fig. \ref{fig:optics} shows the design of the optical beam transmission system for the NASCO Rx. 
\begin{figure} [b]
\begin{center}
\begin{tabular}{c} 
\includegraphics[width=10cm, bb=0 0 1314 716]{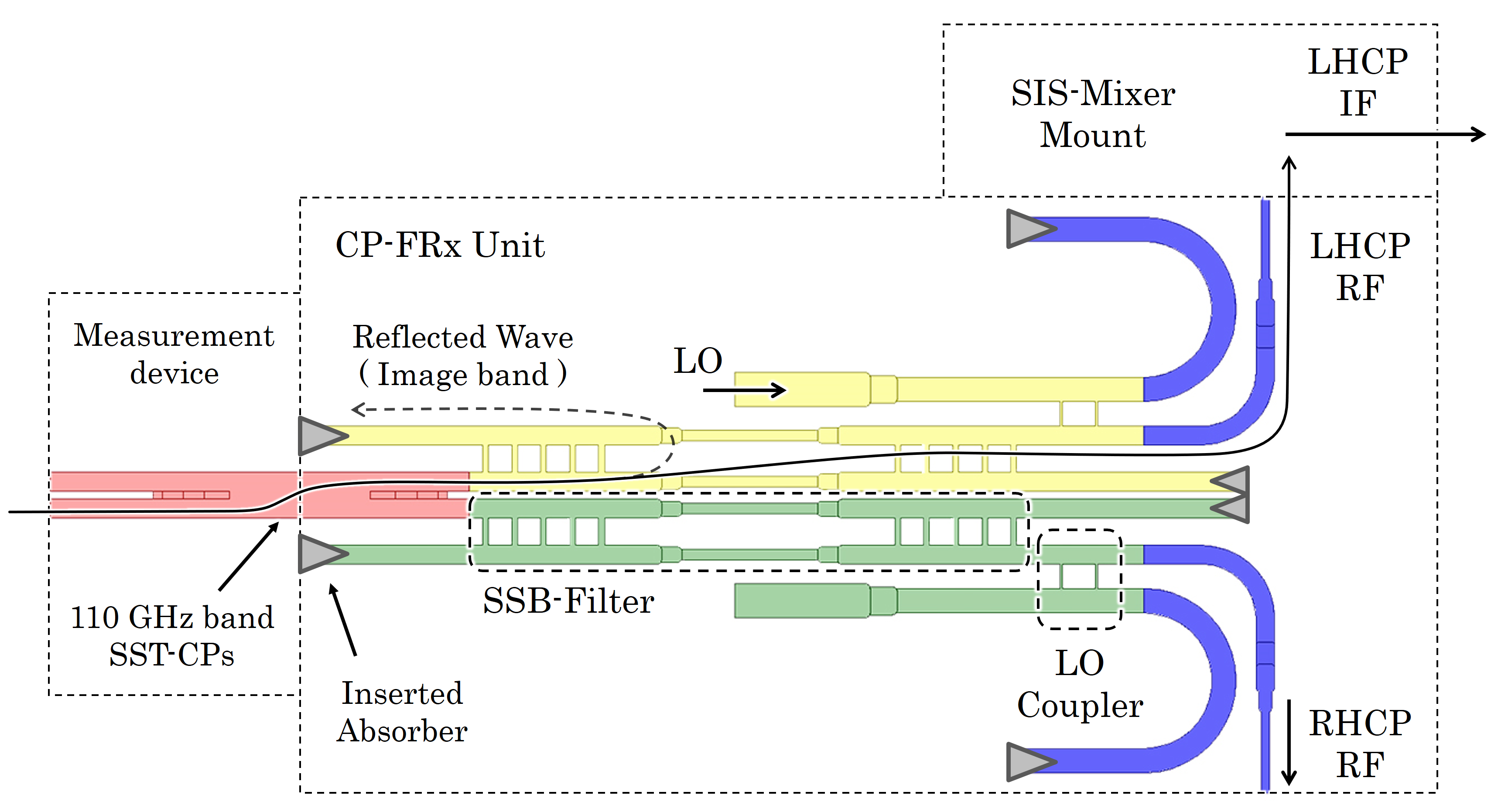}
\end{tabular}
\end{center}
\caption[example] 
{ \label{fig:cpfrx100} 
Schematic of the integrated waveguide circuit inside of the NASCO 110 GHz band CP-FRx Unit.
}
\end{figure} 

\begin{figure} [b]
\begin{center}
\begin{tabular}{c} 
\includegraphics[width=9cm, bb=0 0 917 500]{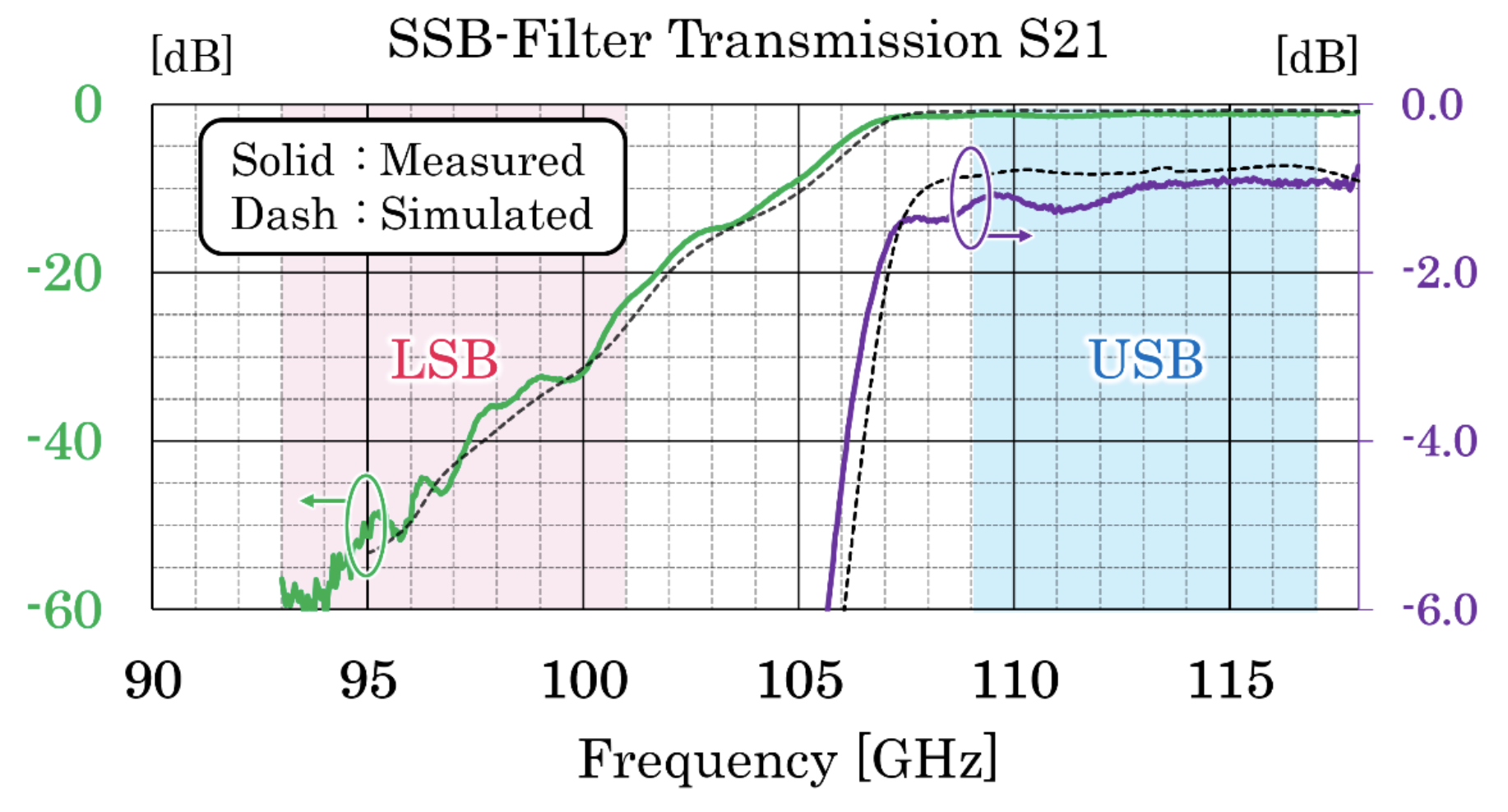}
\end{tabular}
\end{center}
\caption[example] 
{ \label{fig:ssbf-measure} 
Transmission levels (S21) of the SSB-Filter for NASCO 100 GHz band receiver.
Measured and simulated results are indicated as solid and dushed lines, respectively.
The left and right Y axis is only the variation of the scaling, so that the 2 lines colored green and purple show the same performance.
}
\end{figure}

The beams 2--5 (100 GHz band) are focused by the elliptical mirror EM7 \#2--5 respectively, which are equally divided into 90 deg squares with respect to the plane mirror PM6 equipped to outside of the cryostat. 
Then these beams are reflected at PM8 \#2--5 and reach the corrugated waveguide feed horn antenna. 
The beam 1 (230 GHz band) is focused and reflected by the elliptical mirror EM7 \#1 in the direction of 45 deg from the top view to reach the 230 GHz band corrugated feed horn.
In this design, the antenna direction and the axis of the central beam 1 is designed to be matched, while the beams 2--5 have directivity evenly separated from the center. 
The correction for the instrumental error of the telescope driving by the radio wave pointing is performed in the beam 1, and the beams 2--5 is treated as being offset in the Az-El direction.
The pointing errors for the beams 2--5 are corrected for by the analysis software after the observations (see Sec. \ref{sec:pointing}).

\subsection{100 GHz SIS Receiver Front-End with CP-FRx Unit}
\label{sec:rx100}

\begin{figure} [t]
\begin{center}
\begin{tabular}{c} 
\includegraphics[width=11cm, bb=0 0 672 1050]{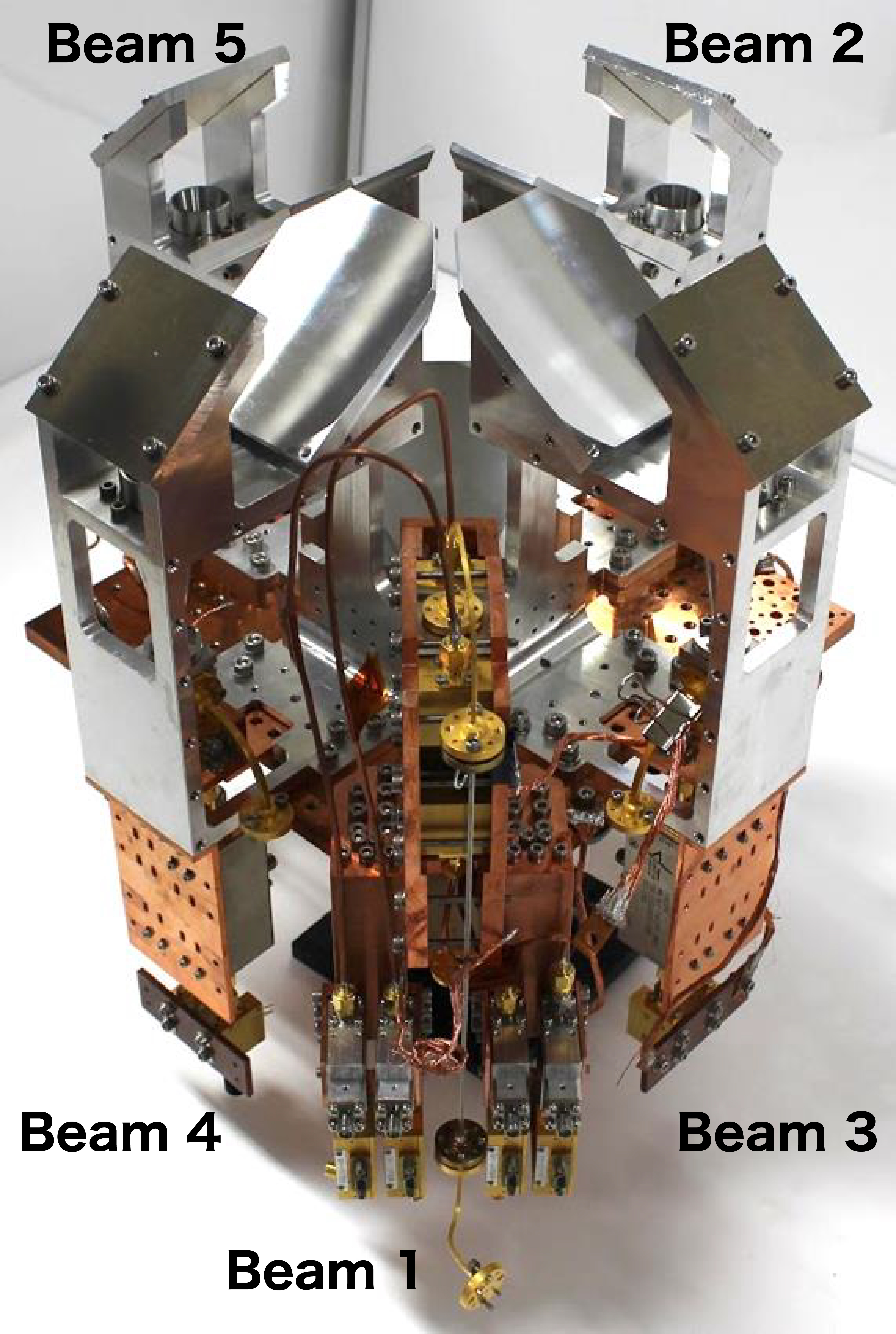}
\end{tabular}
\end{center}
\caption[example] 
{ \label{fig:nasco-rx} 
Photograph of the NASCO receiver. All receivers of 115 GHz band and 230 GHz band are integrated.
}
\end{figure}

Fig. \ref{fig:ifbd100} shows a block diagram of the 110 GHz band receiver for one of the beam 2--5. 
The receiver front-end consists of a corrugated feed horn, a circular polarizer, two waveguide single-sideband (SSB) filters, and two SIS-mixers corresponding to 2-polarization.
The LO frequency of the receiver is 105 GHz, then the RF of 109--116 GHz is output in the IF band of 4--12 GHz.
The circular polarization separation is obtained by a waveguide Stepped Septum Type Circular Polarizer (SST-CP), and SSB separation is provided by an absorption type waveguide high-pass filter. 
These waveguide devices and also LO couplers are compiled into an integrated waveguide circuit device called as the CP-FRx Unit, which is directly connected to the corrugated feed horn antenna.

Fig. \ref{fig:rx100} shows a photograph of the CP-FRx Unit, and the assembled 110 GHz band receiver front-end.
Fig. \ref{fig:cpfrx100} shows the detailed schematics of the CP-FRx Unit internal circuit. 
The CP-FRx Unit has a relatively small square-shaped waveguide input port of $\square$ 1.70 mm, and the input circularly polarized waves are separated from each other with high accuracy by SST-CP, then output them to the two rectangular waveguides. 
This SST-CP is designed as a scaled up and optimized model of the proven higher frequency band SST-CPs reported in Refs. \citenum{2017.JIMRW.Hasegawa, 2017PASJ...69...91H} and \citenum{ 2020.JIMRW.Hasegawa} into the 110 GHz band.

The two absorption type waveguide high-pass frequency filters called as the SSB-Filters are directly connected to the two output waveguides of the SST-CP. 
These SSB-Fitlers are designed as the 110 GHz band optimized model of the ones reported in Refs. \citenum{2017PASJ...69...91H} and \citenum{2015.JIMRW.Asayama}, which consists of the two branch line couplers and the reflected type frequency band-pass filters (BPF). 
For these SSB-Filter designing, the observation RF band of 109--116 GHz including the CO $J=$1--0 emission line is set as the BPF pass band, and thus the rejection band including the RF Image Band of 93--101 GHz is designed to be transmitted forward to the inserted radio-wave absorber by the function of the 90 deg backward coupler\cite{2017PASJ...69...91H}, and then terminated. 
As a result, a high Image-band Rejection Ratio (IRR) exceeding 25 dB is obtained for 109--116 GHz SSB detection.

\begin{figure} [bt]
\begin{center}
\begin{tabular}{c} 
\includegraphics[width=14cm, bb=0 0 866 552]{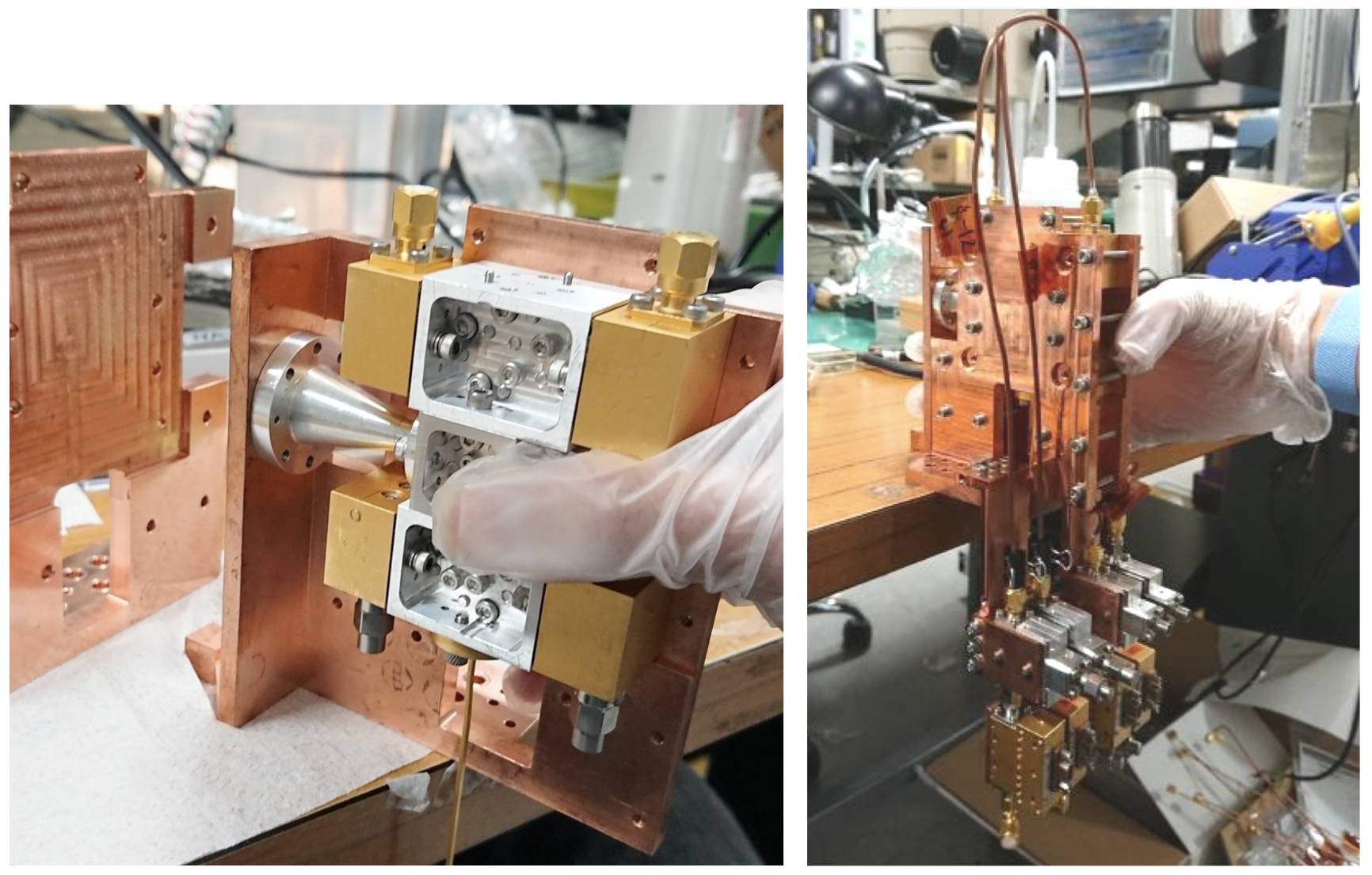}
\end{tabular}
\end{center}
\caption[example] 
{ \label{fig:rx200} 
Photographs of the 230 GHz band 4 K cooled receiver front end.
(Left) The SIS-mixer receiver consists of a corrugated horn, a SST-CP, two 2SB-Filters, and four SIS-mixers.
(Right) The assembled receiver with cooing holders.
}
\end{figure}

Fig. \ref{fig:ssbf-measure} shows the measured transmission level of the fabricated CP-FRx Unit. 
The input RF band in 109--116 GHz is transmitted to the SIS-Mixer with relatively low loss of approximately $-1.0$ dB at 300 K and expected to be $-0.35$ dB at 4 K. 
In addition, the input RF-Image band in 93--101 GHz is absorbed by the SSB-Filter, and only a very small fraction under $-28$ dB at 101 GHz is able to reach the SIS-Mixer, which corresponds to the practical IRR of this SSB-Filter. 
After the output of the SSB-Filter, a $-20$ dB LO coupler and a waveguide size converter to what the SIS-Mixer requires are integrated.
Then, two FOREST 100 GHz SIS-Mixer reported in Ref. \citenum{2019PASJ...71S..17N} is directly connected here. 
The DC bias for these mixers are applied by the 3-port cooling coaxial isolators connected at the IF output ports and the two 4--12 GHz band cooled LNAs, CIT Cryo 4-12A, are connected to the isolator output ports.

Fig. \ref{fig:nasco-rx} shows the photograph of the integrated NASCO receiver.
The receiver is modularized for each beam to improve maintainability.
The module consists of mirrors, a horn, a CP-FRx unit, SIS mixers, and CLNAs.

\subsection{230 GHz SIS Receiver System}

\begin{figure} [tb]
\begin{center}
\begin{tabular}{c} 
\includegraphics[width=15cm, bb=0 0 1366 1000]{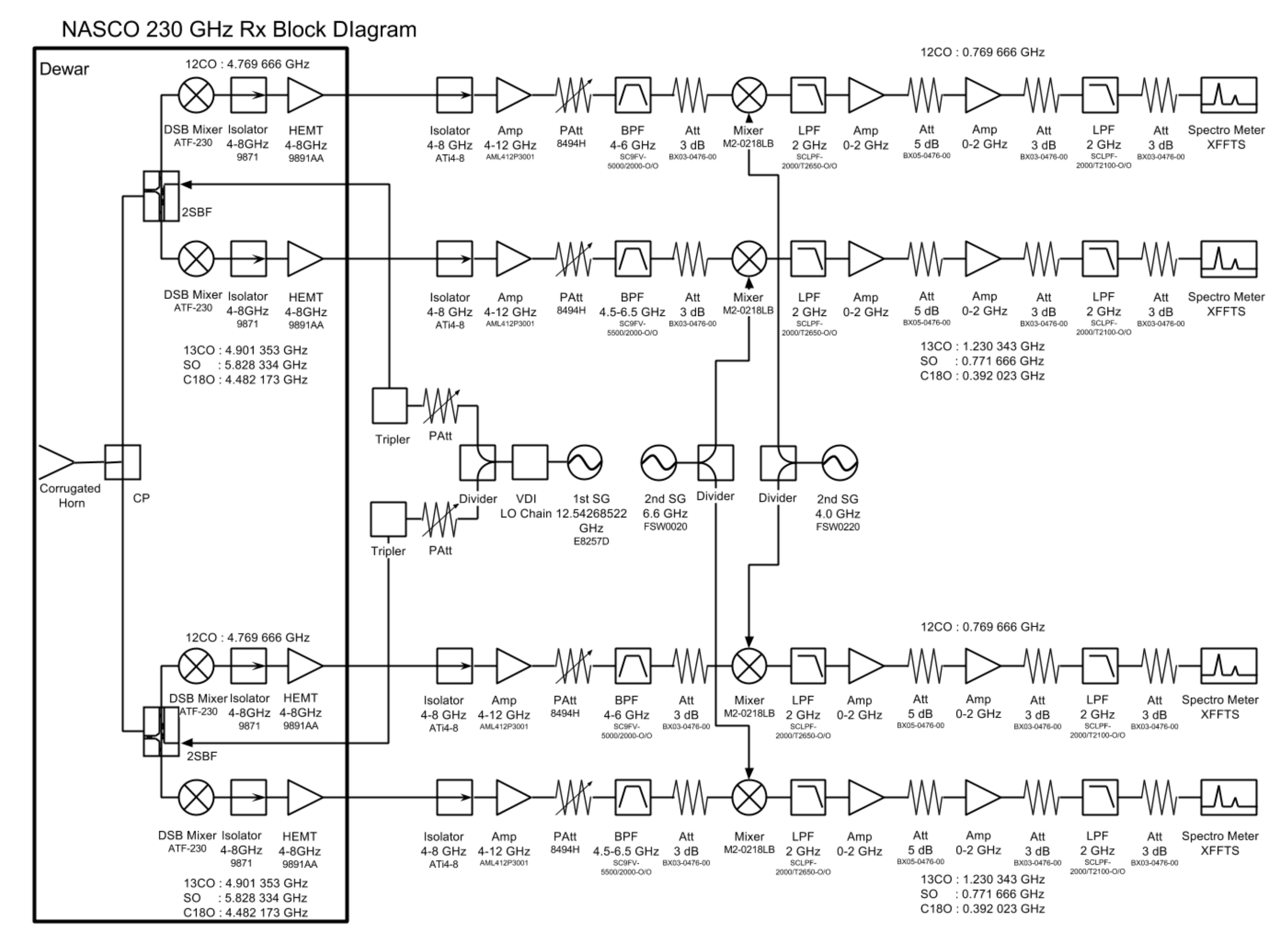}
\end{tabular}
\end{center}
\caption[example] 
{ \label{fig:ifbd200} 
Schematic block diagram of the 230 GHz band receiver system.
}
\end{figure} 

The 220/230 GHz band receiver system is developed as a duplicated one of the observationally proven 220/230 GHz receiver installed in the OPU 1.85-m telescope reported in Refs. \citenum{2017.JIMRW.Hasegawa} and \citenum{2017PASJ...69...91H}, except for an optimized optical interface matched with the NANTEN2 optics. 
Fig. \ref{fig:rx200} shows photographs of the 230 GHz band receiver, and Fig. \ref{fig:ifbd200} shows its block diagram. 
The 4 K cooled receiver front end consists of a 230 GHz band SST-CP and two waveguide frequency band diplexers called as 2SB-Filter, four SIS-Mixer and four IF amplifier chain. 
A 2SB-Filter includes a LO signal divider and two $-20$ dB LO couplers. 
The operational LO frequency was set to 225.5 GHz, and the amplifying IF band was 4--8 GHz, so that the observing USB RF band was 229.5--233.5 GHz including $^{12}$CO $J=$2--1 line, and LSB RF band was 217.5--221.5 GHz including $^{13}$CO and C$^{18}$O $J=$2--1 lines. 
These USB and LSB RFs were separated from each other by the 2SB-Filters with high and stable IRRs over 15 dB.

\subsection{4 K Cooled Cryostat}
\label{sec:cryo}

\begin{figure} [t]
\begin{center}
\begin{tabular}{c} 
\includegraphics[width=16cm, bb=0 0 520 384]{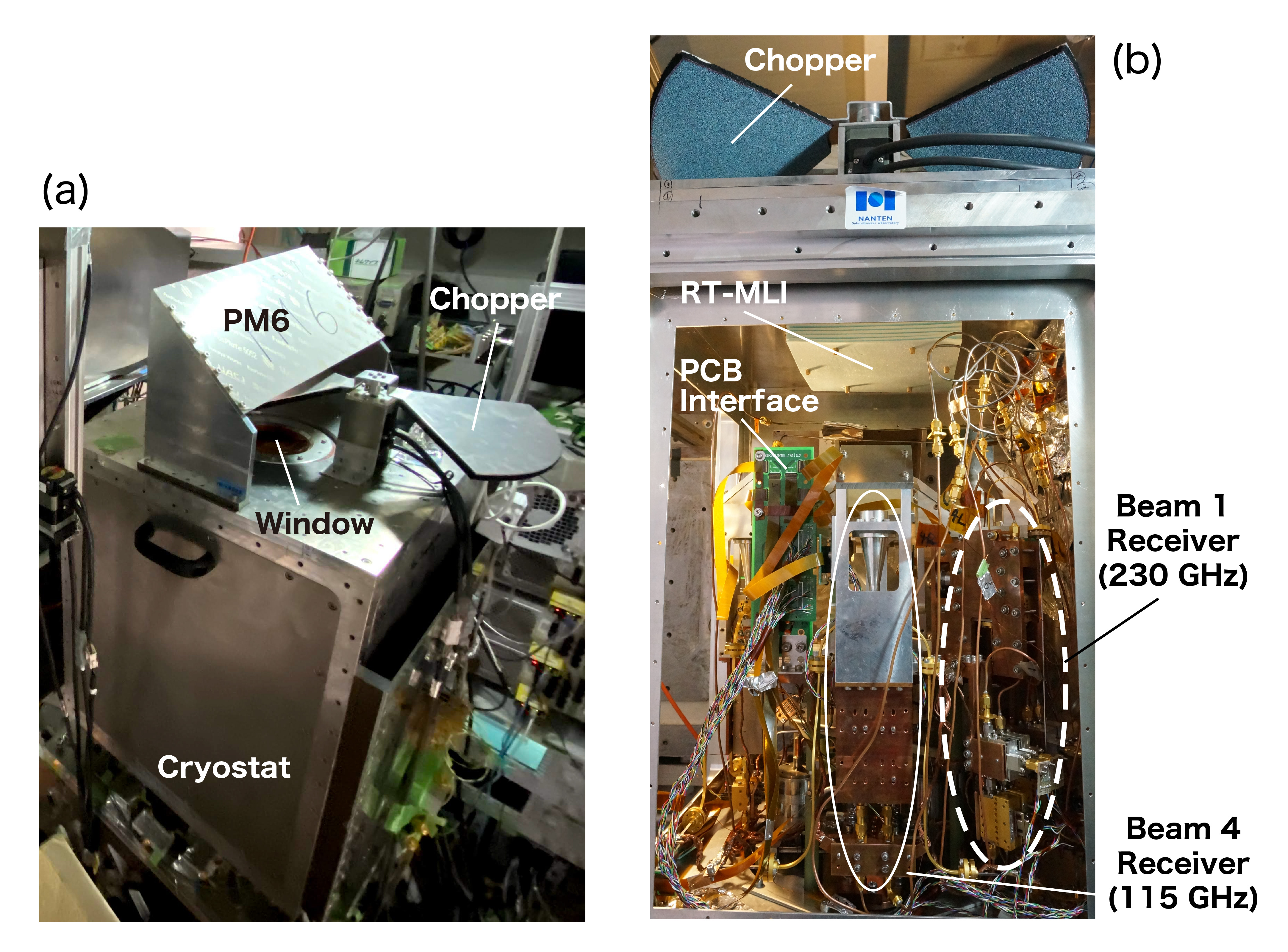}
\end{tabular}
\end{center}
\caption[example] 
{ \label{fig:cryo} 
(a) Photograph of the NASCO 4 K-cooling cryostat installed in the receiver cabin of the NANTEN2 telescope and 
(b) photograph of the inner side of the NASCO Rx. 
}
\end{figure}

Figure \ref{fig:cryo} shows the photographs of the 4 K cryostat installed into the NANTEN2 telescope, and the internal receiver systems.
Because NASCO Rx has receivers for 5 beams with dual polarizations, a large amount of heat inflow from the RF window through thermal radiation and physical connections such as the 12 coaxial cables, 9 LO waveguides, and more than 100 of the bias cables controlling SIS mixers and LNAs as well as heat generation of 12 LNAs are expected.
In order to reduce heat inflow through the RF window, we use the radio-transparent multi-layer insulation (RT-MLI\cite{2013RScI...84k4502C}).
Using a commercial material, Styroace-II Styrofoam provided by the Dow Chemical Company, we made a nine-layer RT-MLI with a thickness of 2 mm to reduce the heat inflow through thermal radiation down to 10\%. 
In addition, we use very thin and integrated flat cables for heat insulation, and then wiring PCB circuits are installed on the 80 K thermal shield stage. 
The estimated value of the total heat inflow of the NASCO Rx is $\sim 1.2$ W.
Using the GM-JT mechanical cryocooler with a cooling capacity of 3.5 W at 4.2 K, the NASCO Rx is cooled stably.

\section{CONTROL SYSTEM: NECST}
\label{sec:necst}

In order to operate the NASCO receiver and carry out the all-sky survey, the telescope control system should be implemented with the capability of handling high data rate (up to 420 Mbps) and fast scan OTF mode.
However, computers and network devices used in NANTEN2 were installed at the same time of the telescope installation of 15 years ago, and it is impossible to achieve above requirements with the old-fashioned system.
Therefore, we replaced all computers and network devices as well as the telescope control software to modern, more powerful and flexible system: the SSD equipped computers and the network switches supporting 5GBASE-T communication.
After the replacement, the networking throughput is measured to be 4.56 Gbps between the computer controlling spectrometers to the storage computers, and it is confirmed that all spectral data are properly saved into the storage.

\begin{figure} [t]
\begin{center}
\begin{tabular}{c} 
\includegraphics[width=10cm, bb=0 0 473 236]{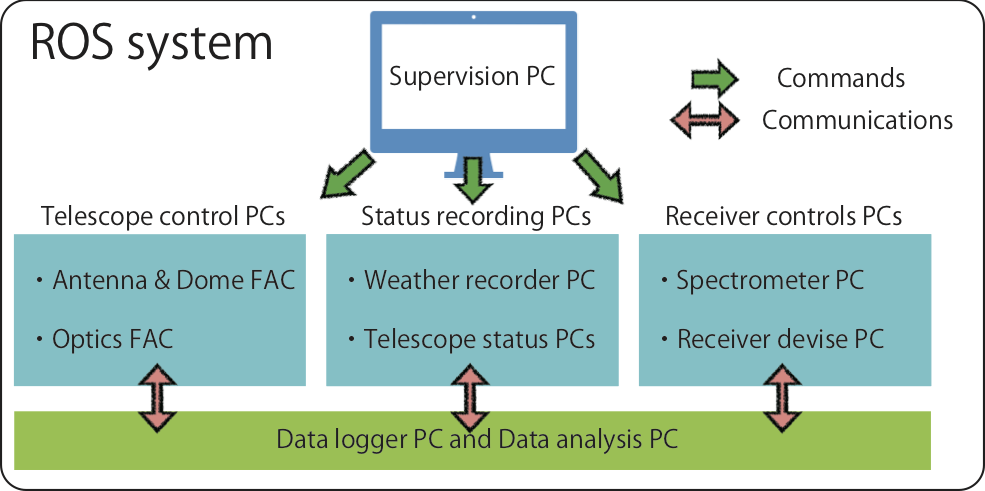}
\end{tabular}
\end{center}
\caption[example] 
{ \label{fig:necst} 
Schematic diagram of the NECST implementation for NANTEN2/NASCO.
}
\end{figure}

In addition, we changed the observation program from C based fully home-made system to Python and ROS\cite{2009.ROS} (Robot Operating System) based system.
The ROS is one of the most prevailing open-source framework to control robots.
The communication module provided by ROS (called Topic), many-to-many communications between processes or computers are easily implemented and maintained.
Using ROS, we developed new telescope control system (named NECST), which is a package of device drivers for the control devices of telescope and receiver, communication modules based on ROS, and some utilities.
More detail for the NECST is described in Ref.~\citenum{2020SPIE.kondo}.
For the NECST implementation of the NANTEN2/NASCO system, totally 104 nodes are running distributed on the 12 computers, and the number of topics which always communicating is 138 (Fig. \ref{fig:necst}).
The scalability and flexibility of the ROS allowed us to build the new control system rapidly.

\section{PERFORMANCE}
\label{sec:csv}

\subsection{System Noise Temperature}
\label{sec:tsys}

\begin{figure} [b]
\begin{center}
\begin{tabular}{c} 
\includegraphics[width=13cm, bb=0 0 1584 936]{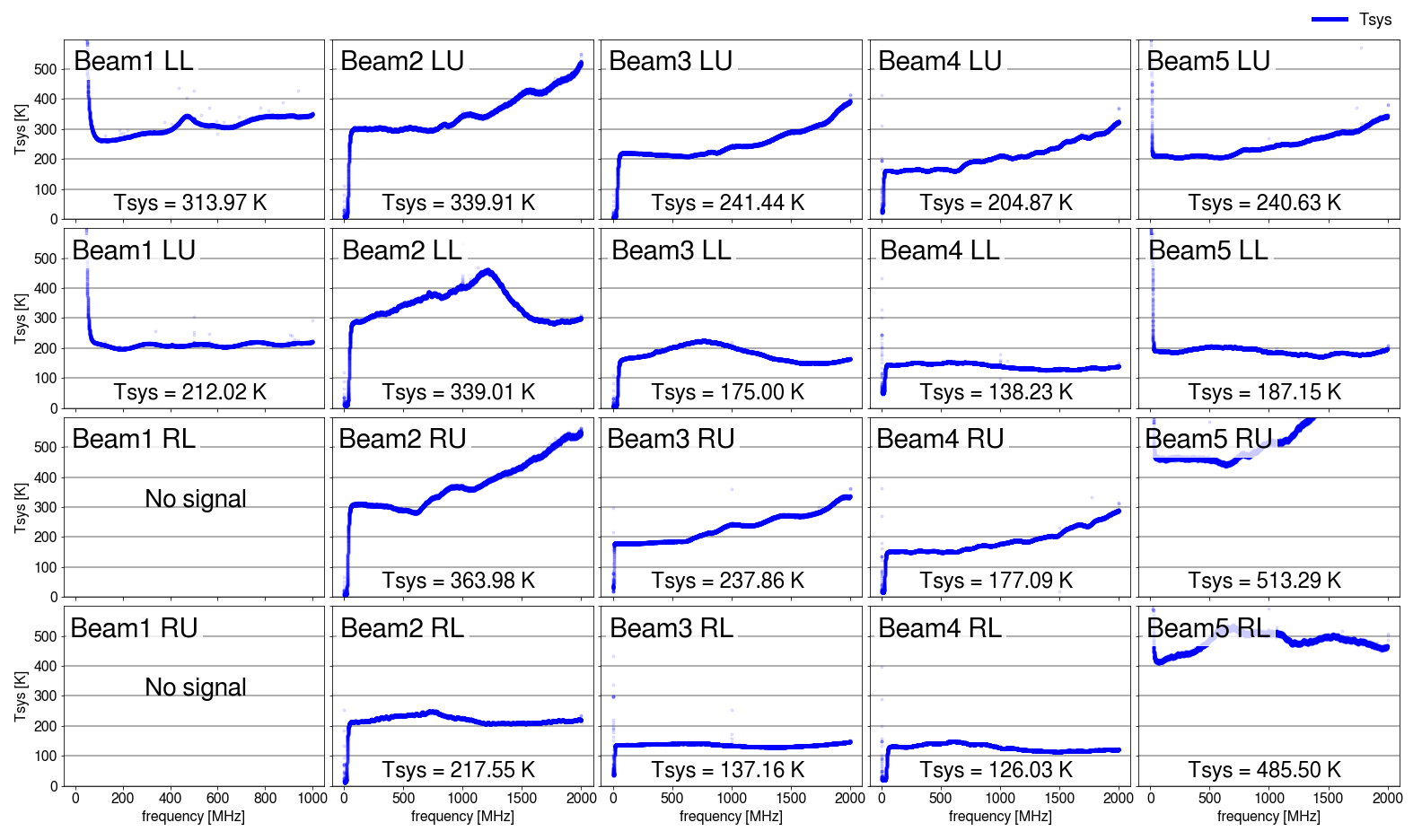}
\end{tabular}
\end{center}
\caption[example] 
{ \label{fig:tsys} 
System noise temperature ($T_{\rm sys}$) of the NASCO Rx on NANTEN2 measured on Mar. 19 2020. 
The direction of the telescope elevation was 80 deg.
}
\end{figure} 

The system noise temperature ($T_{\rm sys}$) including the receiver noise temperature and atmosphere was measured at the Atacama Desert after the installation of the NASCO Rx on the NANTEN2 telescope (Fig. \ref{fig:tsys}). 
The receivers on beam 1 (230 GHz band) of right-hand polarization output no signals, that might be due to malfunction of SIS mixers, and some receivers on 115 GHz band show $T_{\rm sys}>$ 300 K more than twice as the best performance receiver, that might be due to the mulfunctions of SIS mixers and/or LNAs.
For receivers working properly, typical $T_{\rm sys}$ were 120--300 K.
However, these measurements were done in the summer season at Chile (March 19), with a relatively higher humidity condition, and therefore it is not the best performance expected and will be that the performance would be improved in the winter season.
The receivers showing malfunctions should be replaced or repaired, but the COVID-19 situation have been blocking access to the site after the initial performance evaluation described above.
As soon as after the COVID-19 situation is terminated, we plan to fix the receivers and hope that the full performance of the NASCO Rx is obtained.

\subsection{Pointing Accuracy}
\label{sec:pointing}

The beam sizes of the NASCO Rx are 180 arcsec and 90 arcsec for 115 GHz band and 230 GHz band, respectively, and thus the pointing accuracy should be better than 20 arcsec even when the antenna is driven at a velocity of 600 arcsec s$^{-1}$ which is a typical speed of the fast scan mode of the NASCO observations.
We measured the antenna driving accuracy of the NECST implementation for the NANTEN2/NASCO by ordering the telescope target Az-EL coordinates imitating the fast scan of the NASCO observations and by checking output of the encoders.
The antenna driving accuracy was measured to be better than 10 arcsec during the scan with a velocity of 600 arcsec s$^{-1}$.
The installation error of the altazimuth mounting is calibrated by the optical pointing observations, because there are only a limited number of point-like sources for a small-aperture telescope at mm-submm wavelength.
After the optical pointing calibration, the pointing rms error toward the sky coordinate is measured to be about 5 arcsec.

\begin{figure} [t]
\begin{center}
\begin{tabular}{c} 
\includegraphics[width=15cm, bb=0 0 536 238]{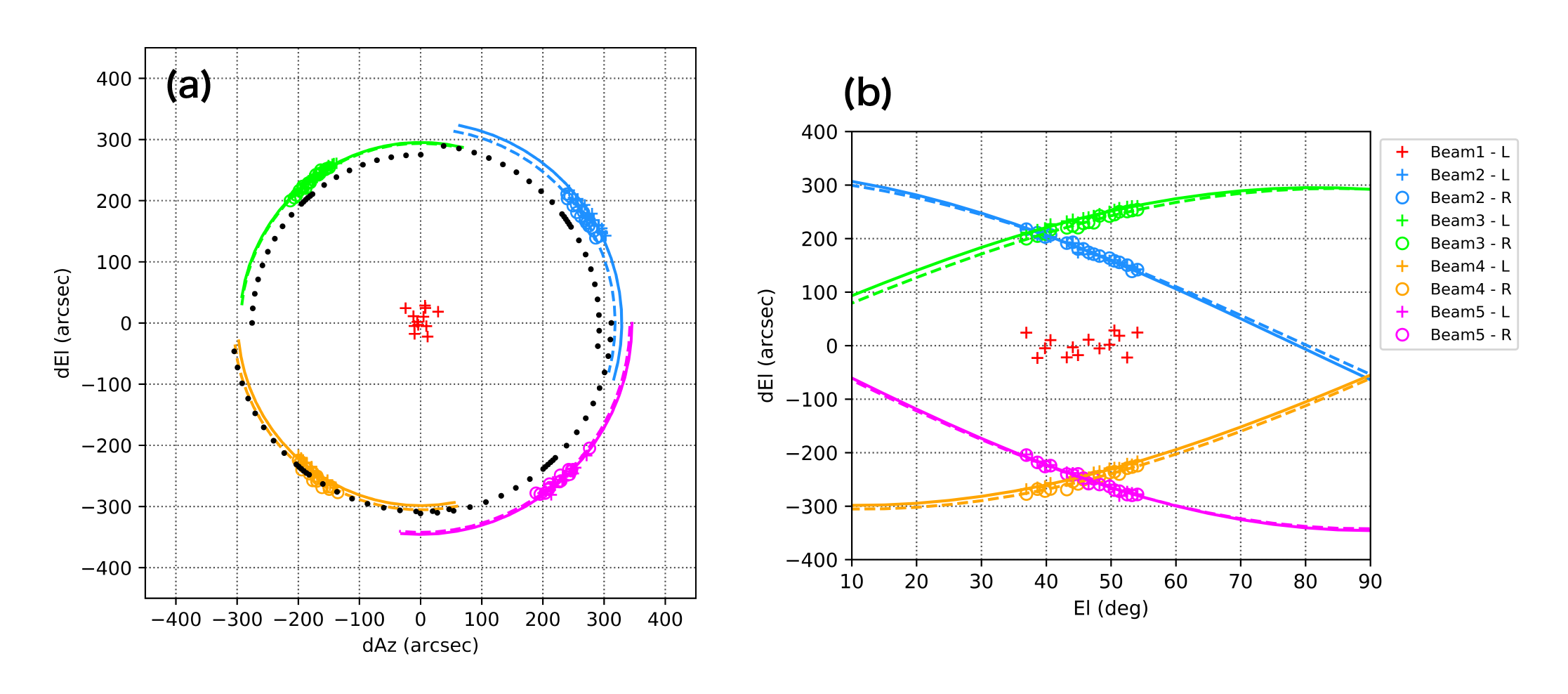}
\end{tabular}
\end{center}
\caption[example] 
{ \label{fig:radio-pointing} 
Results of the calibration observations for radio pointing. 
(a) Scatter plots of the radio pointing residuals. Markers indicate observed value and lines indicate the best fit results of the collimation model for beam 2--5.
Dotted lines show the ideal cases of  pointing offsets for the 100 GHz beams.
(b) Scatter plots of the radio pointing residuals in Elevation-dEl space. 
Markers and lines are same as (a).
}
\end{figure} 

The offset between the optic axis of the beam 1 receiver (230 GHz band) of the NASCO Rx and the light axis of the optical telescope used for pointing calibration was measured by scanning the Sun at various elevation angles, and the measured error was calibrated.
Because a beam rotator is not equipped in the NASCO Rx, the optic axes of the beams 2--5 are rotated with the changes of the elevation angle.
The accuracy of the radio pointing for the beam 1 receiver is measured to be about 5 arcsec (see also Sec. \ref{sec:optics}).
The offset parameters of beam 2--5 receivers toward the optic axis of the beam 1 receiver were also measured by scanning the Sun.
Fig. \ref{fig:radio-pointing} shows the results of the calibration observations for radio pointing performed on 2020 March.
The pointing model for the beam 2--5 can be described as the following equations:
\begin{equation}
dAz = r \cos(\theta - El) + d1
\end{equation}
and
\begin{equation}
dEl = r \sin(\theta - El) + d2,
\end{equation}
where $Az$ and $El$ are the encoder values for the azimuth and elevation angles, respectively; $dAz$ and $dEl$ are the corrections, $d1$ and $d2$ are the offset between the rotation center of the beam and the optic axis, respectively; $r$ is the offset between the beam axis and the rotation center, and $\theta$ is the phase corresponds to the location of the beam axis.
The coverage of the elevation angle was unfortunately relatively limited (35--55 deg), whereas observation results were well fit by the pointing model of beam 2--5, and the pointing parameters of beam 2--5 were obtained.
For the beam 2--5 receivers, the radio pointing accuracy is estimated to be $\sim 5$ arcsec after applying the pointing calibration.
Table \ref{table:pointing} summarizes the best fit parameters obtained for beam 2--5.

\begin{table}[ht]
\caption{Pointing calibration parameters for beam 2--5 of NASCO Rx.} 
\label{table:pointing}
\begin{center}       
\begin{tabular}{ccccc} 
\hline
\rule[-1ex]{0pt}{3.5ex}  Beam & Pol. & $r$ & $\theta$ & Residual error  \\
\rule[-1ex]{0pt}{3.5ex}   &  & (arcsec) & (deg) & (arcsec)  \\
\hline
\rule[-1ex]{0pt}{3.5ex}  2 & L & 329 & 78.8 & 5.2  \\
\rule[-1ex]{0pt}{3.5ex}  2 & R & 318 & 80.3 & 4.2  \\
\rule[-1ex]{0pt}{3.5ex}  3 & L & 295 & 171.6 & 2.9  \\
\rule[-1ex]{0pt}{3.5ex}  3 & R & 294 & 171.6 & 3.7  \\
\rule[-1ex]{0pt}{3.5ex}  4 & L & 299 & $-79.3$ & 5.3  \\
\rule[-1ex]{0pt}{3.5ex}  4 & R & 306 & $-78.4$ & 5.5  \\
\rule[-1ex]{0pt}{3.5ex}  5 & L & 346 & $-0.1$ & 5.2  \\
\rule[-1ex]{0pt}{3.5ex}  5 & R & 342 & $-0.9$ & 5.8 \\
\hline 
\end{tabular}
\begin{tablenotes}
For all beams and polarizations, same parameters of $d1 = -8.0$ arcsec and $d2 = 35.0$ arcsec were used.
\end{tablenotes}
\end{center}
\end{table}

\subsection{Beam Size}
\label{sec:beamsize}

\begin{figure} [b]
\begin{center}
\begin{tabular}{c} 
\includegraphics[width=12cm, bb=0 0 1306 916]{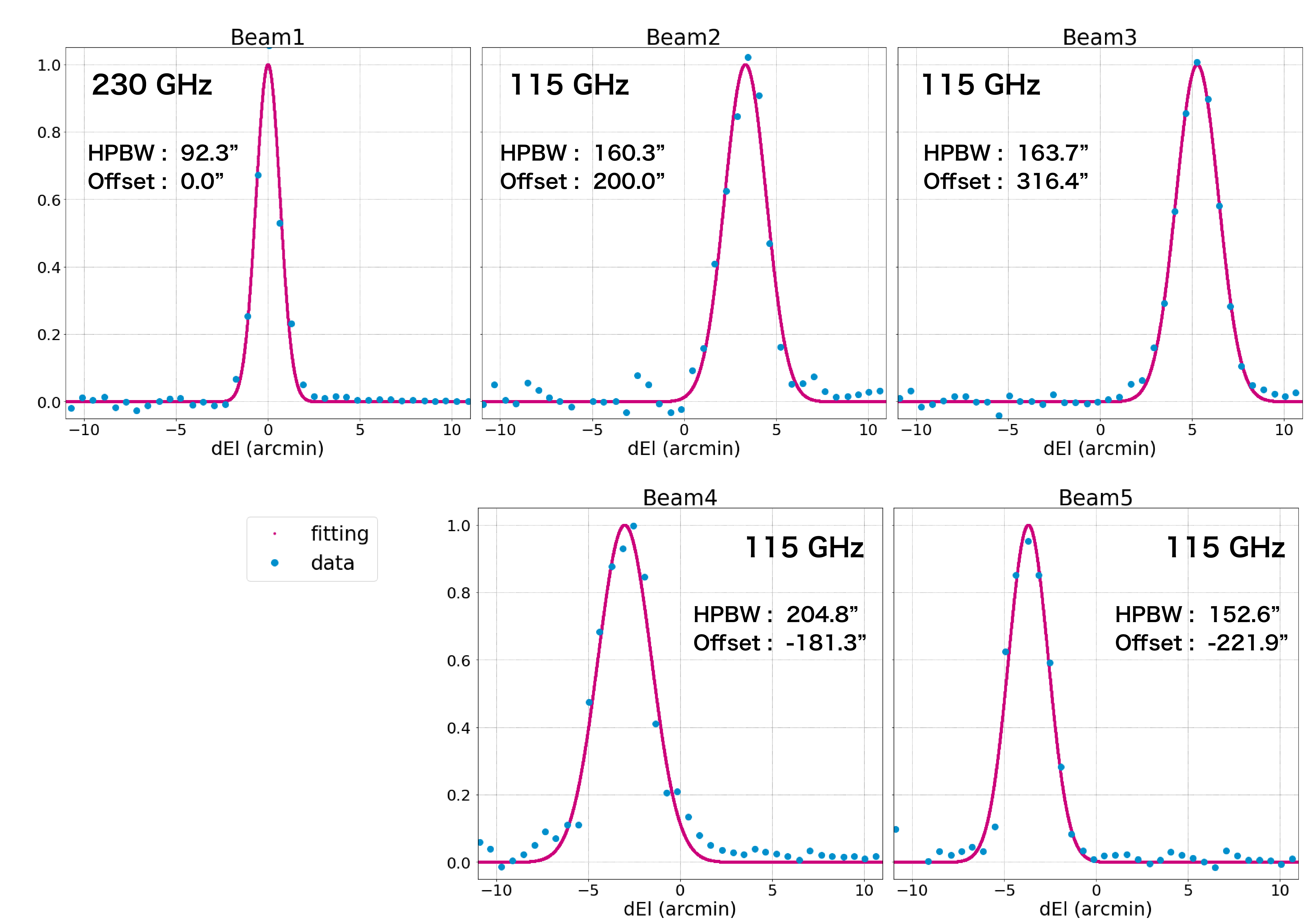}
\end{tabular}
\end{center}
\caption[example] 
{ \label{fig:beam} 
Measured beam pattern obtained as the differential scanning data of the Sun.
}
\end{figure}

The beam pattern was obtained as the differential scanning data of the Sun (Fig. \ref{fig:beam}).
The beam patterns of each polarization were well coincident with each other.
The half power beam widths (HPBWs) of the NANTEN2 telescope with NASCO Rx were measured to be $\sim 90$ arcsec for 230 GHz band and $\sim 160$ arcsec for 115 GHz band except for the beam 4, by assuming a Gaussian-shape beam pattern.

\begin{figure} [H]
\begin{center}
\begin{tabular}{c} 
\includegraphics[width=16cm, bb=0 0 293 374]{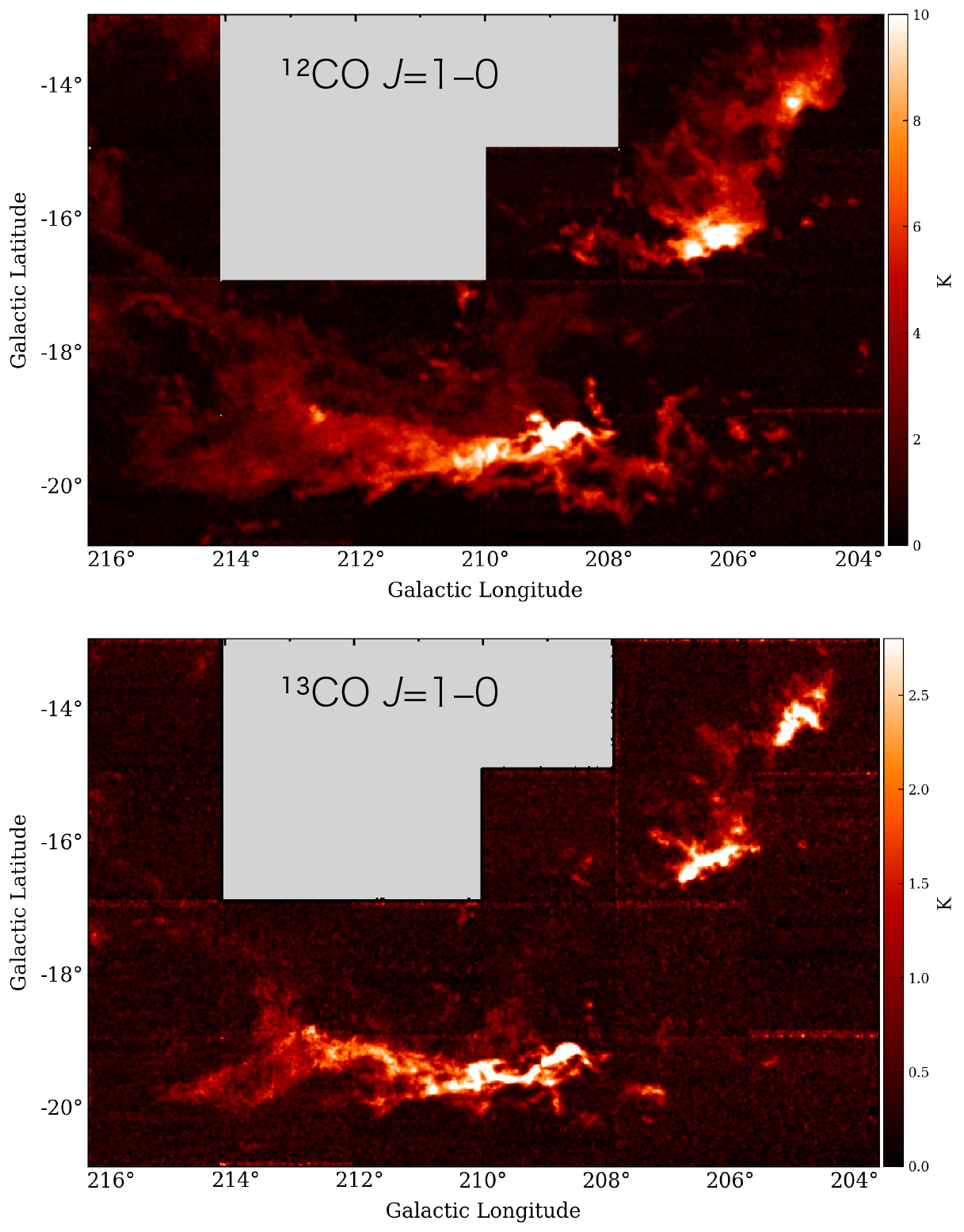}
\end{tabular}
\end{center}
\caption[example] 
{ \label{fig:ori} 
Peak intensity maps of $^{12}$CO and $^{13}$CO $J=$1--0 lines toward the Orion A and B GMCs observed by the NASCO receiver on NANTEN2.
}
\end{figure}

\section{TEST OBSERVATIONS}

Test observations of the $J=$1--0 transition of $^{12}$CO, $^{13}$CO, and C$^{18}$O lines were performed in 2020 March with the fast scan OTF mode toward the Orion A and B GMCs covering the area of 76 deg$^{2}$.
The observation unit of the OTF scan is a $2^{\circ}\times2^{\circ}$ submap with a dumping time of 0.1 s.
The typical observation time of a submap was 80 min, and the total observation time was 25 hours.
Obtained submaps were analyzed and merged by the analyze software package for NASCO.
Fig. \ref{fig:ori} shows the peak intensity maps toward the Orion GMCs.
These demonstrate excellent mapping performance of the NASCO Rx and NECST system.

\section{SUMMARY}
\label{sec:summary} 

We developed the new multi-beam receiver NASCO Rx, which consists of 4-beams of the 100 GHz band receivers for observations of $^{12}$CO, $^{13}$CO, and C$^{18}$O $J=$1--0 and a 1-beam of the 200 GHz band receiver for observations of $^{12}$CO, $^{13}$CO, and C$^{18}$O $J=$2--1.
The observations of the double polarizations are available for both the 100 GHz and 200 GHz band receivers.
We also developed the new telescope control system NECST based on Python and ROS (Robot Operating System) framework, which realizes the handling of the high data-rate operations and the fast scan mapping observations.
The receiver and control system were installed on the NANTEN2 telescope by December 2019, and the performances of the NASCO Rx and NECST system were measured.
Although some malfunctions were found on the SIS mixers and/or CLNAs, the receiver shows effective performance on the large-scale mapping observations.
The typical system noise temperatures ($T_{\rm sys}$) including the atmosphere were measured to be 200--300 K and 120--300 K for the 230 GHz band and the 115 GHz band, respectively, for the elevation angle of 80 deg.
The pointing accuracy was carefully checked and calibrated especially for 100 GHz beams because the pointing errors of those beams are changed with the elevation angle.
The pointing errors for all the beams were measured to be better than $\sim 5$ arcsec.
Using the receiver and control system, the test observations of large-scale CO mapping were performed.
The area of 76 deg$^{2}$ was mapped in 25 hours, demonstrating the excellent mapping performance of the NASCO system.

\acknowledgments 

First of all, the authors would like to honor the commitments by Dr. Akio Ohama who was leading the project especially for the receiver team from the early period on it and have passed away unexpectedly just before the installation of the system.
The project would not have been achieved without his great efforts.
We owe many thanks to the late Ms. Yumi Fujii, who passed away in July 2015. 
She was a superconductor device engineer and made great contributions to this project by providing the low-noise SIS junctions.
The authors would like to thank all people who engaged in this project, especially Taku Nakajima, Sayaka Mochizuki, and Makio Ito.
This work was supported by JSPS KAKENHI Grant Number JP15H05694.

\begin{figure} [h]
\begin{center}
\begin{tabular}{c} 
\includegraphics[width=5cm, bb=0 0 4912 3264]{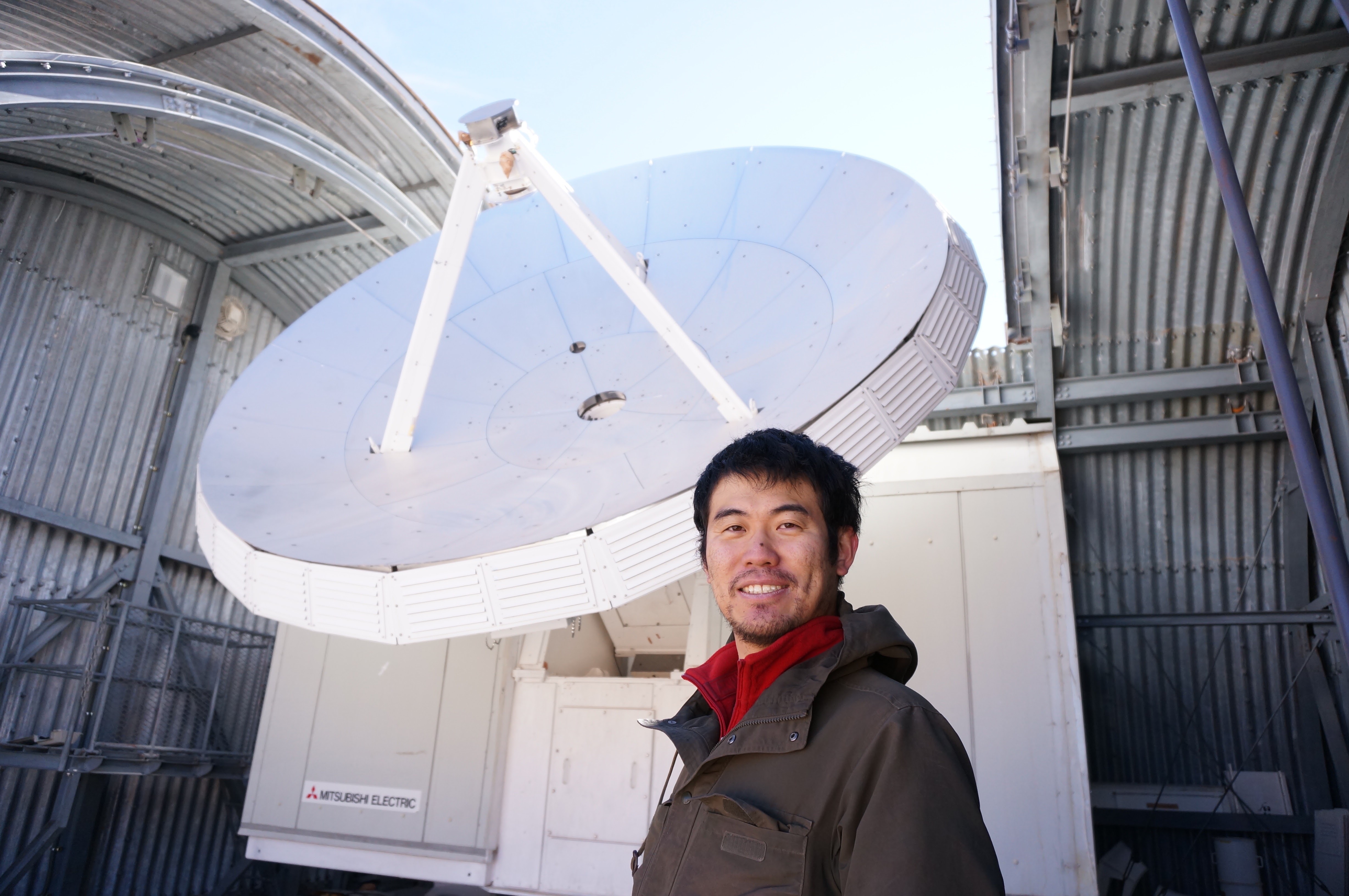}
\end{tabular}
\end{center}
\caption[example] 
{ \label{fig:ao} 
Photograph of Akio Ohama with NANTEN2.
}
\end{figure}

\bibliography{main} 
\bibliographystyle{spiebib} 

\end{document}